# "Make It Sound Like a Lawyer Wrote It": Scenarios of Potential Impacts of Generative AI for Legal Conflict Resolution

Kimon Kieslich[1,2] · Natali Helberger[1,3] · Nicholas Diakopoulos[4]

## Abstract

Generative AI (GenAI) tools are transforming critical societal domains, including the legal sector. While these tools create opportunities such as increased efficiency and potential improvements in access to justice, they also present new challenges, such as the risk of inaccurate legal advice and questions about the legitimacy of legal decisions. However, the full impact remains to be seen and ultimately depends on the way GenAI tools are implemented and used by both, legal professionals and citizens. This makes anticipating and managing the positive and negative impacts of GenAI use in the legal domain challenging but also important to guide the digital transformation of the legal sector into a societally desirable direction. In this paper, we set out to explore the spectrum of possible impacts of GenAI in the legal domain, examining how this technology is anticipated being used and the potential implications this might have for the legal sector and society. Using a scenario writing method, we surveyed participants in the EU and US including both citizens and legal professionals about the potential impact of generative AI on legal conflict resolution. Respondents were tasked with writing a narrative drawing on their experience or expertise about a future in which AI is used throughout the legal process. We qualitatively analysed the prevalence of risk and benefit themes, as well as the types of anticipated legal tasks. We then compared these findings based on expertise status (legal experts versus citizens) and regional regulatory background (the EU with the EU AI Act versus the US with an industry self-regulatory approach). Finally, we describe the emerging trade-offs that will affect decision-makers in the legal sector.

---

✉      Kimon Kieslich (corresponding author)
        k.kieslich@uni-hohenheim.de

[1] AI, Media & Democracy Lab, University of Amsterdam, The Netherlands
[2] UKUDLA, University of Hohenheim, Germany
[3] Institute for Information Law, University of Amsterdam, The Netherlands
[4] Communication Studies & Computer Science, Northwestern University, Evanston, United States

## 1. Introduction

Generative AI (GenAI) is transforming virtually every field of professional activity. The recent launch of Anthropic's legal plugin was a powerful reminder how much this is also true for the legal sector.[1] According to Anthropic, *Claude Legal* must speed up the review of contracts, enable non-disclosure agreement triage, draft legal briefs and change workflows. More generally, in the legal sector, GenAI tools are already being used to draft legal documents and provide legal counsel, support legal interpretations, answer legal questions, predict the outcomes of court cases, and summarize and analyse legal documents (Chien et al., 2024; Chien & Kim, 2024; Choi et al., 2023; Villasenor, 2024). For citizens, GenAI tools can potentially offer a simpler, more comprehensive, and less expensive method of obtaining legal advice. People can use GenAI to draft documents for formal communication with authorities, to receive advice on how to pursue legal conflicts, and to get guidance on employment law and civil lawsuits. Legal professionals are also increasingly using GenAI in their daily work. Surveys of legal professionals have shown that they value the efficiency of these tools, and many intend to continue using them in the future (Chien & Kim, 2024; ELTA, 2023), though the risk of hallucinations and wrongful advice is also real.

Next to numerous potential benefits, the use of GenAI in the legal domain can also present risks to individuals, organizations, institutions, and society. By now notorious is the case of the New York lawyer who was caught using GenAI to produce a briefing full of fabricated case law,[2] and yet other cases followed.[3] Instances of judges using GenAI in the Netherlands were met with public disapproval.[4] Next to the general concerns about the reliability and accuracy of GenAI outputs that apply across sectors, the use of GenAI in the legal domain raises questions on the individual level and how this relates to the procedural rights of citizens, but also at an institutional level (e.g. transformation of the way judges and legal professionals work) and societal level (Dhungel, 2025; Helberger, 2025; Metikos, 2024). Concerns about the impact of AI on the quality and fairness of judicial proceedings are also the reason why the EU AI Act classifies certain uses of AI by judges for the administration of justice as a high-risk application, subject to strict risk monitoring and other legal obligations.

The growing popularity of GenAI in legal practice highlights the importance of being able to identify and anticipate potential risks from the use of GenAI in order to be able to steer the digital transformation of the sector into a direction that is societally acceptable, ethical and desirable (Kieslich et al., 2025; Pullen et al., 2026). (Mandatory) AI risk or impact assessments are a central instrument to do exactly this (Selbst, 2021; Stahl et al., 2023). The objective is to safeguard societal and economic values by intervening as much, but also not more than is needed to make sure technologies are safe. However, many risk assessment methods a) suffer from power imbalances and inclusion deficits, particularly when it comes to incorporating the perspectives

---

of citizens' and affected communities, b) fail to consider potential future pathways and miss potential use cases and their associated risks, and c) are methodologically constrained, focusing nearly exclusively on quantifiable metrics (Kieslich et al., 2025). These shortcomings are especially concerning in the legal sector, where the consequences from unmanaged risks are significant for those involved, and developing a medium-to long-term perspective on how to safely and legitimately transform the sector requires guidance how to make informed decisions about risk-benefit trade-offs.

In this study, we examine the use of GenAI in legal conflict resolution as a case study to reveal which risks citizens and legal professionals anticipate and how they navigate trade-offs. We define a legal conflict as a dispute between two or more parties that involves a disagreement over legal rights, obligations, or interpretations of the law. Importantly, GenAI could be used in all stages of legal conflict resolution such as in formulating a complaint, information gathering, analysis of the law, document drafting, and even judicial decision-making. For many citizens, legal conflicts are tangible. According to an EU report, there is an average of at least one incoming civil, commercial, administrative, or other legal case per 100 inhabitants in each EU member state per year. There are also considerable differences between member states, with some countries, such as Denmark (~40/100 inhabitants) and Austria (~35/100 inhabitants), showing a particularly high number of legal cases (European Commission. Directorate General for Justice and Consumers, 2025). In the US roughly two thirds of the population are involved in some kind of legal conflict over a four year time span (The Hague Institute for Innovation of Law & The Institute for the Advancement of the American Legal System, 2021).

Empirically, we present findings from a scenario-writing study focusing on the future of generative AI (GenAI) in legal conflict resolution. Scenario writing is a well-established research method in future studies and anticipatory governance. It uses storytelling grounded in socially and technically plausible events to identify technology development trajectories and tap into the lived experiences of the narrative composers (Amer et al., 2013; Barnett et al., 2025; Börjeson et al., 2006) and is a viable tool to engage in ethical debates and legal reasoning (Helberger, 2024). We sampled scenarios from (i) citizens and (ii) legal professionals in the EU and US (for each group n=25, total N=100). We then analyse which risks and benefits are anticipated and the purposes for which GenAI might be used. Finally, we compare the anticipatory perceptions of legal professionals with those of citizens and across regulatory frameworks (EU vs. US): While the EU introduced a unified legal framework with the EU AI Act that mandates risk assessment for (high risk) AI systems for the European market, the US follows an industry self-regulatory approach. Against this background it is relevant to compare the perceptions of EU and US legal professionals and citizens as they are indicative for the regulatory reach for both regions. Our qualitative analysis inductively identifies themes in the written narratives. Consequently, our study identifies helps to uncover potential future debates and areas of research in risk management, policy development, and evaluation within the legal domain.

## 2. Risk Management of GenAI in the Legal Sector

A common approach to safeguarding the implementation of new AI tools (including GenAI) in society is to use risk-based approaches to address the challenges and opportunities of AI. These approaches require risk and impact assessments and management. The EU AI Act proposes a framework that classifies AI systems into four categories: unacceptable risk, high risk, limited risk, and minimal risk. Some instances of using GenAI in the legal domain fall under the AI Act's high-risk category. According to its Annex 3, para 8a, AI used in the "Administration of justice and democratic processes" constitutes one of the high-risk use cases for AI. According to the AI Act, this includes systems "used in researching and interpreting facts and applying the law to concrete facts or used in alternative dispute resolution" (Recital 61 AI Act). Additionally, certain General Purpose AI models can be classified as having "systemic risks", depending on their size and potential impact on society (Article 51 AI Act). The classification as 'high risk' or a model with 'systemic risk' brings with it a legal obligation to engage in risk management. Other instances of using AI in the legal area do not fall under the high-risk category, such as when lawyers or citizens use AI. This raises the question of whether the use of AI by citizens or other legal professionals besides judges is likely to pose lower risks for society. Having adequate risk assessment methods in place is important not only for compliance with existing laws such as the AI Act, but also for identifying potential gaps in protection.

In comparison, the US is lacking a federal legislated framework for AI governance. Former president Biden introduced Executive Order 14110 (Executive Order on Safe, Secure, and Trustworthy Development and Use of Artificial Intelligence) that laid out propositions for risk management of AI technologies. Yet, this order was revoked by president Trump in early 2025. Current risk management of AI technology in the US is primarily based on voluntary commitments to using frameworks or standards (e.g. from ISO), with some requirements also emerging from state legislatures (Nazareno & Douglas-Glenn, 2025). The National Institute for Standards and Technology (NIST) offers its AI Risk Management Framework[5] which provides suggestions for technology companies on how to identify and mitigate risks caused by AI systems (National Institute of Standards and Technology, 2023, 2024). But again, the application of this framework is entirely voluntary and does not explicitly mention the legal sector.

In general, researchers across Europe and the US have critiqued existing approaches to risk assessment (Efroni, 2021; Griffin, 2024; Orwat et al., 2024). They have highlighted that the practical implementation of risk assessments is typically expert-focused and often put in the hands of the providers themselves. This can lead to companies engaging in limited practices of identifying only those risks they identified initially and that are comparatively easy to address, which leads to negligence of more structural risks, such as inequalities and inconvenient outliers (Griffin, 2024). Neither under the NIST framework or the AI Act are there any legally binding requirements to involve non-expert stakeholders, such as citizens or vulnerable groups, in risk assessment processes or decision-making over risk acceptance or management (Ebers, 2024; Ullstein et al., 2025), which puts companies in the driver's seat to decide on value conflicts in risk management (T. Gillespie, 2024). However, perceptions and lived experiences offer valuable insights that could significantly enhance the identification and evaluation of AI tools and challenge the "objective" nature of traditional risk assessment methods (Fischer-Abaigar et al., 2024; van der Heijden, 2021). In fact, risk assessment does not occur in a "neutral" vacuum but involves highly contextual decisions impacted by cultural, individual, societal, and political factors that require human judgment (Orwat et al., 2024; Schmitz et al., 2024). Legal conflicts, as a case in point, are arguably greatly influenced by individuals' lived experiences. Identified impacts may vary greatly among affected individuals depending on available financial resources, knowledge of the legal system, and access to justice.

Another critique raised by scholars is the fixation of current risk assessment standards and regulatory approaches on already known risks, i.e. well-established risks that can easily be measured (Ebers, 2024). However, there are also risks that a) may only materialize in the future

---

and b) risks that are entirely unknown thus far but that could have a significant impact on individual and social life. These risks are inherently more difficult to measure, and current practices are largely insufficient because they focus heavily on quantifiable risks while largely ignoring qualitative and exploratory approaches (Orwat et al., 2024). The recent introduction of GenAI in the legal sector as a case in point has led to the emergence of many new impacts, which are still being explored. These impacts are influenced by both the technical capabilities of the tools and how users – including legal experts and citizens – will utilize these tools for legal purposes. Put differently: whether or not certain risks materialises depends not only on the development and design of the technology itself, but also on the way it is being deployed and used by professional users (lawyers, clerks, judges) and other users such as citizens. Understanding the possible implications and ways in which these users anticipate using GenAI-assisted legal tools can help developers, regulators and decision makers to develop a more forward looking vision on desirable and less desirable futures, potential factors of influence and intervention points.

Lastly, current risk assessment practices focus on the quantitative measurement of risks, i.e., making risk mapping scalable. However, scholars argue that valuable contextual information is lost when risks are reduced to established metrics (Gellert, 2020; Pasquale, 2023; Walker et al., 2024) – information that contains important viewpoints such as societal, individual, and political context. Subsequent trade-off decisions in risk management are often based on quasi-objective calculations of risk scores rather than more qualitative, deliberative processes (Stahl et al., 2023). We agree with this observation and highlight the legal domain as a critical example. Whether or not the use of GenAI in the legal sector is societally acceptable and contributes to the realisation of public values (such as access to justice, fairness, etc.) is the result of weighting risks and benefits for a concrete solution (such as legal conflict resolution), the way affected users prioritise values and negotiate risks.

Thus, building on our previous work (Kieslich, Diakopoulos, et al., 2024; Kieslich et al., 2025; Kieslich, Helberger, et al., 2024), we apply an alternative method for conducting risk assessments and management for high-risk use cases using written scenarios drafted by diverse stakeholder groups to identify and evaluate GenAI risks. In collecting and analysing written scenarios from different stakeholder groups, we are able to ground the current debate on impact of legal GenAI tools in lived experiences and prospective outlooks from different angles, including perspectives that are commonly neglected in risk assessment practices. We especially highlight the socio-technical nature of risk and benefit assessment and are also able to identify plausible trade-off applications for the use of GenAI. In this study, we focus on a GenAI application domain where the impact may be particularly severe for individuals and society: the legal sector.

## 3. The Current and Projected Use of GenAI in the Legal Sector

Generative AI (GenAI) is already being used for various tasks in the legal sector, including legal research, administrative work, preparing drafts, and supporting legal decision-making (Laptev & Feyzrakhmanova, 2024). According to the 2023 Legal Professionals & Generative AI Global Survey commissioned by the European Legal Tech Association, 44 percent of the respondents indicated weekly GenAI use. A majority of respondents (77%) also indicated that GenAI will help them in the future (ELTA, 2023). Additionally, a 2024 survey revealed that 41 of the 100 largest U.S. law firms currently utilize AI (Henry, 2024). Legal professionals identify access to legal information, preparing summaries of documents, and document drafting as applications, while they express concerns about data security, algorithmic bias, and accountability (ELTA, 2023).

In a field study in which researchers provided legal aid professionals with one month of access to generative AI tools, 90 percent of respondents reported increased productivity, and 75% intended to continue using GenAI tools for work in the future (Chien & Kim, 2024). They identified opportunities in document summarization, confirmatory preliminary research, producing first drafts, and translation tasks (Chien & Kim, 2024). These efficiency gains are supported by Schwarcz et al. (2025) who reported significant productivity benefits for legal practitioners who use GenAI in their work. Similarly, another study reports that GenAI tools can increase citizens access to the court system by helping translate information into other languages or more accessible language, curating legal provider information, and providing guidance through self-help forms (Chien et al., 2024). An interview study of German judges regarding the acceptability of AI use in the legal sector revealed that judges view the automation of routine tasks (e.g., issuing orders) positively, yet they are more critical of using AI to calculate risk scores for individuals and cases (Dhungel & Beute, 2024). In a similar vein, a case study utilizing a sample from the Brazilian Supreme Court reported that benefits of AI were seen in decreasing the time needed to derive legal judgments and help shifting activities of legal professional to more complex tasks (De Sousa et al., 2022). In short, proponents of GenAI in the legal sector often cite efficiency and cost savings, as well as increased access to justice, as the main reasons to move legal conflict resolution to the digital sphere (Helberger, 2025; Socol De La Osa & Remolina, 2024).

Despite their potential, researchers question the quality of legal advice produced by GenAI tools. Magesh et al. tested the reliability of the three legal research AI tools: LexisNexis, Thomas Reuters and GPT-4. They found that even the best performing specialized legal AI tool produced hallucinations in 17% of queries (Magesh et al., 2024). Non-expert tools, such as GPT-4, performed substantially worse, producing inaccurate legal advice in 43% of queries. The authors list several reasons for these inaccurate results: failing to retrieve the most relevant sources (naïve retrieval); a mismatch between the cited documents and the actual query (inapplicable authority); the tendency to agree with the user (sycophancy); and reasoning errors (Magesh et al., 2024). Additionally, legal professionals are hesitant about the potential impact of AI on their profession. Based on a survey of Portuguese judges, Martinho (2024) reports that risks are seen in the lack of a human element in decision-making, the support in drafting legal documents, and in the erosion of legitimacy of the legal profession as a whole. Helberger (2025) corroborates these findings by analysing the challenges for the legal sector at the institutional level. She points out that the rise of AI tools presents several fundamental challenges to the legal sector. These include the materiality of institutions (shifting toward online conflict resolution rather than physical buildings, e.g., courts); the jurisdiction of courts, which is affected by technology companies' global outlook when creating legal tools; and the erosion of legal expertise and states' monopoly on adjudication, facilitated by the emergence of digital, AI-powered tools that are offered by commercial technology companies. These developments challenge "expertise, authority and claims over a certain jurisdiction" (Helberger, 2025, p. 9). This could lead to an erosion of trust in judicial systems, as AI decisions might amplify societal biases and reduce the explainability of court decisions (Socol De La Osa & Remolina, 2024). Thus, it is evident that the use of AI in the legal sector comes with value trade-offs, both at the level of individual cases and at a more fundamental level between technology companies and the judiciary.

Previous studies have shown that the public is rather sceptical about the use of AI in the legal sector, particularly for automated judicial decision-making (*AlgoSoc AI Opinion Monitor*, 2025; Brauner et al., 2025; Kieslich et al., 2021, 2026; Kim & Peng, 2024). However, these studies focus broadly on acceptance ratings for the entire sector and fail to delve into detailed projections on how generative AI (GenAI) might impact legal conflict resolution in the future. For example, the prospections of citizens are largely absent from current empirical studies (Chien & Kim, 2024; Dhungel & Beute, 2024; ELTA, 2023). Furthermore, little is known about the motivations and anticipated habitualisation of using GenAI for legal conflict resolution, as well as the cultural, professional, and societal implications of its use. Though, this information is crucial for defining safety standards for AI tools used in the legal sector and developing guidance for users of the technology (Socol De La Osa & Remolina, 2024).

In this paper, we investigate which anticipated legal tasks (Section 5.1) are plausibly be impacted by GenAI. Further, we map risk (Section 5.2) and benefit (Section 5.3.) perceptions for the use of GenAI for legal conflict resolution. For each of these dimensions, we provide contextualized examples that narrativizes the prospections of the different stakeholders. In doing so, we add tangible insights that can enrich the debate about risk management for GenAI tools in the legal sector. Further, we compare anticipations between legal professionals and citizens and across the two judiciaries EU and the US (Section 5.4). To get a better insights into trade-off perceptions in regard to different stages of the legal process, we also compare risk and benefit anticipations for the use of GenAI for different legal tasks (Section 5.5).

## 4. Method

### 4.1. Approach

To uncover plausible risks, benefits, and purposes of AI tools in their contextual embedding, we conducted a scenario-writing study. Scenario-writing is an established research method in future studies and anticipatory governance that uses storytelling to uncover future trajectories of technology development (Amer et al., 2013; Börjeson et al., 2006; Ramirez & Wilkinson, 2016). Written scenarios are not predictions about the future, but rather show plausible developments based on the current state of the world. Scenarios are especially well-suited to activate a wide spectrum of stakeholders, including citizens (Barnett et al., 2025; Ramirez & Wilkinson, 2016). This study builds on previous work by the authors who adapted and refined scenario writing to study the impacts of GenAI (Kieslich et al., 2025). In our application of the method, we emphasize the socio-technical element, i.e., the interplay of humans with technology.

### 4.2. Procedure

Empirically, we conducted an online survey among citizens as well as legal professionals in the EU and the US using the survey pool of Prolific[6]. Field time was from March 24 to April 15, 2025. For each group we sample 25 respondents totalling 100 respondents. We are interested in the EU in the context of the common legal AI regulation framework (EU AI Act), which establishes regulations for all EU member states. We study the EU as a whole due to the limited availability of legal experts that can be reached via Prolific. To avoid comparing a national sample of citizens with an EU sample of legal experts, we also defined EU citizens as the focus. In contrast, the US follows a largely unregulated approach on the federal level and relies more on industry self-regulation. Including both EU and the US respondents in the sample, allows us to compare differences and similarities across regulatory approaches.

---

The survey was administered using Qualtrics in the English language, although respondents were offered the possibility to write their scenario in their native language[7]. In these cases, we used DeepL to translate the written scenario to English. Automatic translation has been proven to be a reliable tool in communication research (De Vries et al., 2018; Lind et al., 2022). After informing participants about the study and obtaining their informed consent, respondents were presented with the instructions for the scenario-writing exercise (see Appendix A). They were instructed to write a fictional scenario exploring how generative AI technology could affect legal conflict resolution five years from now. To activate the respondents' lived experiences, they were asked to base their stories on their professional or personal perspectives, experiences, and knowledge and to anticipate a situation in the near future. After receiving more information on the basic capabilities of GenAI technology (see Appendix B), respondents were asked to confirm that they understood the instructions and that they would not use LLMs to draft the stories. To ensure an adequate amount of creative thinking and prevent speeding, respondents were required to spend at least 25 minutes drafting their scenarios. After submitting their scenario, respondents were asked to reflect on the potential risks they identified in the use of GenAI as described in their scenario. Finally, participants had the opportunity to provide feedback and were debriefed. The average response time for both EU samples was 50 minutes whereas the US respondents needed 43 minutes (Law Professionals) and 48 minutes respectively (Citizens). Respondents were paid 10 EUR[8] respectively for their participation. Furthermore, 10 percent of the respondents received a 3 EUR bonus payment for exceptionally well-written stories. The ethics committee of the first author's institution approved the study.

### 4.3. Sample

To avoid including LLM-generated data in our sample, we screened every scenario for AI-generated language using GPTZero (Tian & Cui, 2023). We contacted all scenario writers with a detection probability greater than 80% and asked them to retract their participation if the assessment was correct. This resulted in the retraction of one scenario in the EU legal professional sample, three in the EU citizen sample, six in the US legal professional sample, and nine in the US citizen sample. After the retractions, we collected enough valid scenarios to reach our target sample size of 25 per stakeholder group. Table 1 shows the sample statistics.

| | EU | | US | |
| --- | --- | --- | --- | --- |
| | Citizens (n=25) | Legal Professionals (n=25) | Citizens (n=25) | Legal Professionals (n-25) |
| Gender | Women = 12 Men = 13 | Women = 12 Men = 13 | Women = 12 Men = 13 | Women = 12 Men = 13 |
| Average Age | 30 years | 34 years | 38 years | 40 years |
| Country of Residence | Italy, Poland, Portugal = 5 Spain = 4 Greece = 3 Hungary, Ireland, Slovenia = 1 | Portugal = 6 France, Germany = 4 Belgium, Greece, Italy, Spain = 2 Estonia, Hungary, Poland = 1 | US = 25 | US = 25 |

*Table 1: Sample Statistics*

## 5. Findings

To analyse the scenarios, we conducted a qualitative open coding analysis using Atlas.ti. Specifically, the first author used qualitative thematic analysis and axial coding (Glaser & Strauss, 2017; Lofland et al., 2022) to uncover risks, benefits, and AI associated legal tasks within the scenarios. The author team, then, discussed the codes in multiple iterations to determine the reported code structure. Four example scenarios (one from each participant sub-sample) can be found in Appendix C, and quotes from other scenarios are included throughout the manuscript. Scenarios written by legal professionals are marked with *LP*, scenarios from citizens are marked with a *C*. Scenarios from the US sample are indicated through an *US*, scenarios from the EU with an *EU*. The number indicates the scenario number from our dataset.

### 5.1. Types of Legal Tasks for Legal Conflict Resolution

Respondents described a wide range of legal tasks (see Table 2). We clustered these tasks along the different stages of a legal process.

*Client interaction & support* includes activities where GenAI-powered chatbots provide case-related information. These are initial consultations that inform citizens (i.e., the affected parties) about their rights as the following scenario describes: "Mark returned home and opened a common generative AI he would use to ask questions to. He began to ask who would be at fault in the scenario he had just experienced, and after some back and forth with adding what context he could remember, was told that in that case he would not be at fault" [US_C_04]. These consultations were also described as a pre-step before consulting a human lawyers "a client, before giving us an assignment, submitted his question to ChatGPT and provided us, together with the documentation, with the prompt and the feedback obtained from the AI." [EU_LP_10]. However, in some scenarios, it was also described how legal professionals used GenAI to facilitate communication with clients. Often, GenAI support was combined with other legal functions, as this scenario describes: "After uploading all their documents and additional information, the AI lawyer informs Angela and Luca that they have a very strong case and predicts with strong confidence that they would win an eventual case, but it still recommends going through with a human lawyer. It still provides an actionable plan and script for their case when asked to, so that the couple can decide to represent themselves for the case or at least have everything prepared to meet a lawyer." [EU_C_13]

*Case management and administration* are tasks typically performed by legal professionals. The goal is usually to reduce the administrative burden by automating tasks such as assigning cases to lawyers ("Opening his computer, he sees the email LawIA has sent him. Every morning, the AI system used by the firm automatically assigns cases to employees." [EU_LP_09]), or translating or transcribing documents ("The software starts uploading her documents and analysing her written declaration of what happened to her in her country and also translated this since it was originally in Spanish" [US_LP_20]). In this case, the GenAI tools have limited autonomy and perform support tasks.

GenAI was also described as taking on *legal research & analysis* tasks. The scenarios depicted legal professionals *and* citizens using GenAI to research case law and find relevant legal information for their cases such as: "Clara uploads all the documents of the case to EUROXY: emails, contracts and the dismissal paperwork. In a matter of minutes, the AI reviews thousands of similar sentences and delivers a full report, pointing out similar cases and proposing ideas for the defense." [EU_C_20]. Furthermore, GenAI was anticipated to be used by legal professionals and citizens for creating claims by summarizing key arguments for a legal process, visualizing evidence, and simplifying language as this scenario describes: "Lawbridge instantly preforms a deep dive into relevant case law both for intellectual property infringement and trademark law. The AI gathers and synthesizes thousands of documents to generate a personalized report for John" [US_C_14]". Additionally, it was also envisioned that GenAI tools could help in also analysing the counter position as this except demonstrates: "Further, generative AI could help find legal research in support of each side's position under Bermuda law - it could also consider if other jurisdictions may be relevant to look to for additional support. This research would be based on the information that the generative AI summarized and it could also find research in support of the other side's position so that it can prepare counter-arguments." [US_LP_12]

GenAI was further described as a tool for gathering *evidence* and for *investigation* purposes. In contrast to legal research tasks, these tools were intended for actual evidence collection, such as recreating a crime or tracing documents as this scenario describes: "there is now the possibility of using AI to create another witness statement. To do this, the AI takes all possible existing evidence and can create its own testimony based on the audio recordings, images of the accident site and vehicles, and witness statements, and can precisely reconstruct the behavior of the accident using a video file, for example, and then indicate whether the accident could have happened as the witness described it." [EU_LP_18]. On the other hand, it was also described that GenAI could help in analysing evidence, for instance, going through uploaded documents, videos and case files to analyse them for legal relevance: "He asks AI to summarize the 20 hours' worth of body camera footage. AI produces a summary of the video footage." [US_LP_03]

A wide variety of anticipated legal tasks provided support in *document and content generation.* Scenarios described citizens *and* legal professionals using GenAI to draft various documents, from simple letters to legal documents and contracts, for instance: "Within very less minutes, it had drafted a formal complaint that sounded more professional than what we could have ever written. It also gave us a fair decisions suggested to the landlord related to many more similar cases across the state" [US_C_24]. These tasks vary considerably in potential impact due to the binding nature of some tasks. Furthermore, GenAI was anticipated to facilitate brainstorming activities where users could discuss legal conflicts and search for solutions.

Lastly, *decision making and adjudication* were also anticipated to be impacted by GenAI. As with document and content generation, the degree of automation varied considerably between scenarios and depending on the level of human input. For example, some scenarios envisioned GenAI playing a supporting role by providing judges and juries with case overviews and real-time references to other legal cases while retaining the decision-making authority with legal professionals. It was also described that GenAI could be used to make a pre-trial assessment: "The firm wanted to know the chance of success if the case went to trial. To get an idea of the likelihood of success, the firm fed case information into a local generative AI tool that was trained on publicly available trial results of motor vehicle accident cases in Illinois. The tool analysed the facts of Doe's case against the backdrop of data of other state cases, and noted a 50/50 likelihood of success." [US_LP_25]. Further, some scenarios envisioned GenAI's potential use in legal mediation, where these tools would provide an overview of the legal situation for both parties and propose a settlement ("Maria and Alexios decide to use a new generative AI legal tool called "Lawyerless" under the supervision of a judge. After inputting all the legal files, as well the whole documentation of their lawyers negotiations, the AI comes up with a solution." [EU_LP_15]). Other scenarios described the fully automated use of GenAI for making automated judgments or acting as AI judges that issue binding rulings: "The judicial side of things [...] have taken quite the turn, once held up in court and in the presence of judge and jury now can be done via your smart phone, you just submit the grievance, create a text explaining the situation and you shall be presented with an AI lawyer, trained on extensive court cases and resolution, advising based solely on success rate and presenting you with bullets points to proceed in order to maximise your chances of coming out on top and even on how to maximize returns on said order." [EU_C_03].

| Main Code | Sub-Codes |
|---|---|
| Case Management & Administration | Administrative Work; Case Assignment; Strategizing; Transcription; Translation; Verification |
| Client Interaction & Support | Chatbot; Consultation |
| Decision Making & Adjudication | AI Arbitration; AI Judge; AI Jury; AI Lawyer; Automated Judgement; Mediation; Pre-Trial Case Assessment; Support in Judicial Decision-Making |
| Document & Content Generation | Brainstorming; Drafting Contracts; Drafting Documents; Drafting Letter; Refining Documents |
| Evidence & Investigation | Analysis of Evidence; Creation of Deepfakes; Evidence Collection; Fact-Checking; Forensics |
| Legal Research & Analysis | Legal Research; Scenario Generation; Simplification of Language; Summarization / Overview of Arguments; Visualization |

*Table 2: Type of Legal Tasks*

## 5.2. Risks

Risks were frequently mentioned in the scenarios. We identified five main categories of risks that encompass a wide variety of subcodes. To align our coding with the AI risk assessment literature, we aligned part of our risk enumeration with the AI Risk Repository (Slattery et al., 2024). The AI Risk Repository is a comprehensive, continuously updated database for categorizing the risks of AI systems. Drawing on 65 other databases and risk frameworks, it provides a high-level overview of risk domains (Slattery et al., 2024). We used the main risk categories of the AI Risk Repository as the main codes for our risk mapping; however, the identified sub-codes in this study are mainly original and only partly overlap with the sub-categories of the AI Risk Repository. Thus, our risk mapping provides novel insights on specific risks, and adds an illustrative note to risk factor identification, while orienting itself to a relatable framework. The identified risks are shown in Table 3.

*AI system safety, failures & limitations* risks are referring mostly to inaccurate results and AI-related errors as this scenario shows: "The main attorney goes to review Ms. Sanchez original declaration and package written by her and contrasts this with the brief that was sent. The AI had summarized and in turn diminished the gravity of the situation and what had actually occurred. Additionally, while translating the documents the software had misconstrued certain meanings of what was originally said which changed the scenario in many ways." [US_LP_20]. Another prominent failure source of legal tools are related to mistakes in judgments, or "hallucinating" legal cases as the basis for its recommendations ("It was inevitable that DINIS would get it wrong and it turned out that one of the legal references he cited… was completely invented" [EU_C_10]). Limitations also commonly include the inherent lack of human qualities associated with AI, such as the inability to comprehend motives, moral or display human intuition as this excerpt exemplifies: "The truth is that although the artificial intelligence can determine who is at fault or not, taking into account the parameters established by law, it doesn't have the sensitivity to understand and determine the amount of compensation because it can't understand or quantify what moral damages are." [EU_LP_03]. Another limitation is the individuality of many legal conflicts, which are described as too complex to be satisfactorily solved by AI tools. Additionally, scenarios also discussed the input dependency of GenAI, i.e. that the quality and comprehensiveness of output is critically dependent on the input of its users, which often warrants existing legal knowledge.

Connected to AI system safety, failures & limitations, *justice-related risks and biases* were frequently mentioned. In general, the scenarios described an increase in the injustice of legal conflict resolution as a result of GenAI use and were mostly connected to individuals. Similar codes describing justice risks were a lack of understanding of right and wrong or the wrongful punishment of parties involved in a conflict. Furthermore, various forms of bias fall under this category, including the system's partiality, lack of context specificity, and lack of cultural understanding: "The plaintiff contended that the suggestions being made by CivAI were in fact biased and disproportionately discriminated against non-native English speakers since the training datasets had been influenced by enforcement patterns that had not been neutral" [US_LP_29]. Most of these risks were described as disadvantaging people with fewer resources or vulnerable populations: "It was also discovered that the platform was trained with internal documents from the big Portuguese firms that helped develop it and this meant that the tool was unintentionally biased in terms of having better legal strategies for cases where clients had more resources, making clients with fewer resources have quicker decisions and less legal assistance so that they were no longer a "dead weight" for the firms or a potential expense." [EU_C_10]

*Governance* risks subsume, in contrast to justice-related risks and biases, systemic risks. For example, the lack of accountability is a concern because it will be more difficult to assign responsibility to specific groups of people if AI is used throughout the legal process ("Who's really responsible for the mistake? Should the developers of the AI tool's be held accountable for allowing a faulty legal document be generated, or is it not their fault?" [EU_C_08]). Furthermore, a lack of transparency refers to the "black box" nature of AI systems, which makes it difficult to detect biases or judge the legal knowledge base of these tools: "Amanda's office was criticized for utilizing unvetted AI-generated outputs and for neglecting to ensure that such outputs were carefully reviewed by human experts, an egregious disregard for the state's Administrative Procedure Fairness Act recently amended to include requirements designed to improve algorithmic transparency." [US_LP_29]. Another concern is the interdependence of AI tools, whereby legal tools may propose different solutions to resolve conflicts and/or perpetuate errors made by other tools as this scenario describes: "Gustave then discovers that LawIA tried to cover up the mistakes of RadarIA. He discovers that the two AIs are covering for each other." [EU_LP_09]. Additionally, societal acceptance of AI refers to the questions of public legitimization of AI questioning the validity of AI use and trust in the results due to public or professional concerns.

*Socioeconomic and environmental harms* primarily pose risks to the legal profession. Scenarios point to the potential devaluation of human legal expertise in light of the accessibility of legal chatbots as describe in the following scenario" "[M]any lawyers, like Joana, have seen their reputations and years of dedication to the field of law tarnished by relying on non-human technology" [EU_C_10]. This is also connected to a potential loss of prestige or reputation for lawyers, either because AI offers a viable alternative or because legal professionals misuse GenAI tools: "At the end of it all, it was fitting that my downfall occurred as quickly as it took AI to do my work for me. The next day, my desk was cleared off." [US_LP_01]. Additionally, there is a concern that legal professionals might lose their jobs and law companies lose revenue because of an uptake of GenAI use in the field. Environmental harms, such as energy costs, play only a minor role in this category.

*Relational risks* mostly concern the role of human involvement in legal conflict resolution in the age of AI. Scenarios suggest that human connections may be lost when citizens consult legal tools instead of talking to the other party involved in a conflict: "I also think it's a risk to eliminate the places where colleagues can meet and develop a dialogue that facilitates a conciliatory solution even outside of mediation or assisted negotiations" [EU_LP_13]. Furthermore, scenarios depict the potential consequences of over-reliance on legal AI tools, which could lead to adverse outcomes for the parties involved (e.g., wrongful punishment): "Since the paralegal had trusted the AI software to give the main ideas she had not noticed what was missing from the original declaration and caused her clients asylum to be denied" [US_LP_20]. From a more philosophical perspective, some scenarios discuss the potential loss of originality and meaning, i.e., humans not knowing how to navigate conflicts and relying on standardized AI outputs.

| Main Code | Sub-Codes |
|---|---|
| AI system safety, failures & limitations | Complexity; False Attributions; Incompleteness; Inconsistency; Input Dependency; Lack of Human Intuition; Lack of Human Understanding; Lack of Information; Low Accuracy; Manipulation; Mistakes; Outdatedness of Training Data; Privacy; Shallowness / Vagueness |
| Governance Risks | Deception; Dependence on AI; Interdependence of AI Tools; Lack of Accountability; Non-Transparency; Power Asymmetry; Social Acceptance of AI |
| Justice-Related Risks & Bias | Alignment with User Perspective; Bias/Partiality; Granularity and Context Specificity; Increase in Injustice; Lack of Cultural Understanding; Lack of; Understanding of Right & Wrong; Wrongful Punishment |
| Relational Risks | Lack of Bridge Building; Loss of Human Interaction; Loss of Meaning; Loss of Originality; Overtrust |
| Socioeconomic & environmental harms | Devaluation of Expertise; Energy Costs; Job Loss; Loss of Reputation; Loss of Revenue |

*Table 3: Typology of Anticipated Risks*

### 5.3. Benefits

Respondents often describe the benefits of using AI for legal conflict resolution. However, as shown in Table 4, the anticipated benefits are less diverse.

Scenarios frequently mention lower or no *costs* as a benefit. These cost benefits materialize for citizens in the form of chatbots ("unable to afford services of a highly profiled lawyer, Rudolf relies on legalbot-499, a generative AI to assist with legal research, draft pleadings and even predict case outcomes based on precedents." [EU_LP_01]), and for law firms that require fewer personnel to fulfil tasks: "It would save time by helping streamlining repetitive tasks like document review, legal research and even help in drafting contracts thus reducing costs for both lawyers and clients." [US_C_24]. Furthermore, the court system is expected to incur lower costs as many cases are anticipated to be resolved before reaching court.

| Main Code | Sub-Codes |
|---|---|
| Costs | *No subcodes* |
| Efficiency & Productivity | Administrative Work; Ease of Use; Efficiency; Spark Ideas; Stronger Client Relations; Time Savings |
| Fairness & Trust | Access to Justice; Impartiality; Transparency |
| Quality of Output | Accuracy; Avoiding Courts; Compelling Language; Comprehensiveness; Personalization; Reliability; Strong Data Base; Testing Arguments |

*Table 4: Typology of Anticipated Benefits*

Cost benefits are often associated with increased *efficiency and productivity*. These benefits mostly consist of time savings, as legal consultations, for example, can be provided almost immediately: "Thank to the AI the legal research concerning different jurisdictions and the draft of a legal opinion and of a letter in a foreign language took just minutes, instead of days of work, helping saving both money and precious time." [EU_LP_02]. For legal professionals, efficiency is associated with reducing the administrative workload and research requirements, and freeing up more time for personal interactions with clients as this scenarios describes: "AI has been doing a good job of uploading documents, taking payments, and even interacting with the clients when they come into the office and the paralegals are busy. There have been no complaints. They have also used this software to draft briefs which helps the paralegals as they are now able to spend less time on this and deal with clients one on one" [US_LP_20]

Scenarios also referred to benefits regarding the *quality* of the legal tools. In these cases, the accuracy and reliability of AI decisions were viewed as positive factors – in contrast to other scenarios that anticipated worsened accuracy and reliability through GenAI for different tasks. Such quality benefits were often associated with anticipated wins of court cases as this scenario shows: "He runs it through AI repeatedly until AI cannot find any loopholes or arguments against it anymore and he is absolutely confident to get his client found not guilty by the jury." [US_C_13] Furthermore, AI is expected to improve the comprehensiveness of legal advice and/or decisions: "With these arguments, Manuel was able to defend himself in court against the entity, without even needing a lawyer, as the AI generated a concise and structured argument that won him the lawsuit." [EU_LP_24]. Quality benefits are also anticipated as GenAI is anticipated to provide compelling language for legal arguments, the potential for personalisation of legal tasks (e.g. personalised legal advice) and the belief in a comprehensive legal database that GenAI relies on. One scenarios describes the benefits as such: "The platform is quite remarkable, and it is able to though the evidence that has been submitted step by step, cross-reference already existing case law, and give users personalized legal advice or some sort of settlements within a matter of minutes." [US_LP_29]" Another benefit was seen as scenario-writers expected less need of going to physical courts as disputes can be resolved easier before going to court with the help of GenAI.

*Fairness and trust* benefits are connected to scenarios that assign AI the ability to make impartial decisions. AI is viewed as a neutral, transparent tool that leads to fairer conclusions ("Sometimes conflicts between people are impossible to resolve, but an impartial AI comes to the rescue." [EU_C_28].). This category also included the anticipation of an increase in access to justice as legal AI tools are cheaper than getting advice from human lawyers: "It is actually here to make justice more reachable to everyone, especially for people like Martin who would not have had the time or money to fight back." [US_C_24]

### 5.4. Differences between expert groups and regional contexts

Next, we conducted an exploratory analysis of the differences between scenarios written by the four different sample groups namely 1) legal professionals from the EU, 2) EU citizens, 3) legal professionals from the US, and 4) US citizens. Due to the huge variety of sub-codes and associated low frequencies for many of them, we decided to analyse difference on the main code level. Note that scenarios can contain multiple categories and that some codes overlap conceptually (this applies to every coding theme). Thus, many scenarios describe the use of GenAI for multiple legal tasks, and associated risks and benefits. For the co-occurrence analysis on the main level, we aggregated the sub-codes into main codes. Hereby, we counted the occurrence of main codes on the scenario level, i.e. if the code was mentioned or not. If scenarios contained multiple sub-codes for one main category (e.g. it mentioned the benefits time savings and support in administrative work) the main code level was still counted as 1. Thus, we avoid inflation of sub-code occurrence. Consequently for each sample group, the maximum occurrence of a main code can be 25. Due to the qualitative nature of this study, these findings should be interpreted as trends hinting at different understandings of GenAI's potential impact on legal conflict resolution.

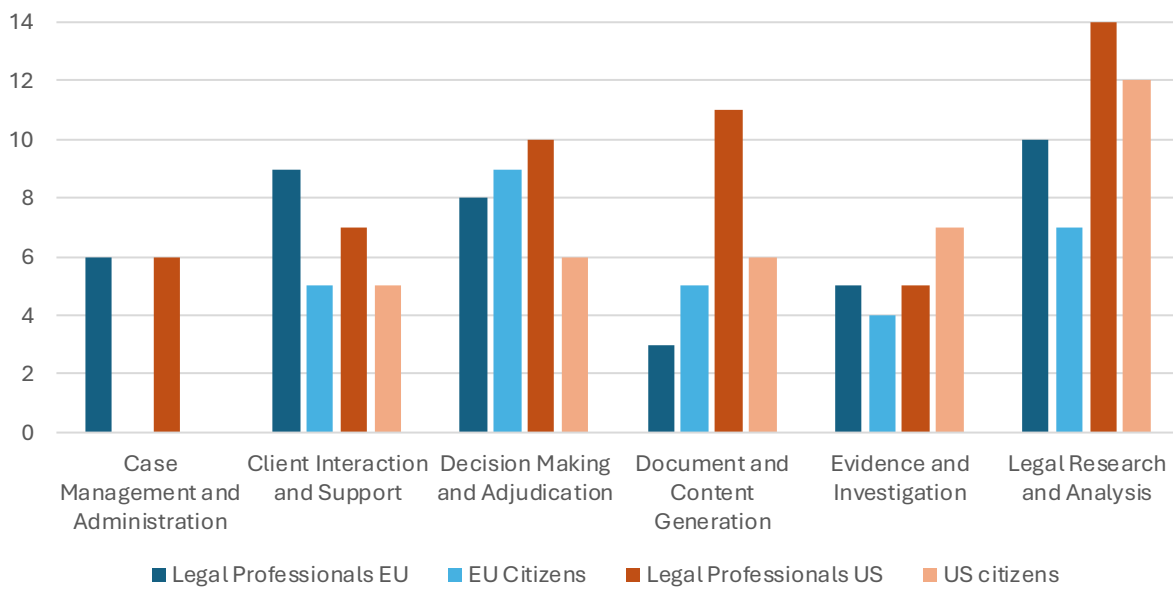

*Figure 1: Prevalence of Legal Tasks per Sample Group*

Examining the prevalence of various legal AI functions reveals clear differences between the sample groups (Figure 1). In general, legal experts (US=53; EU=41) describe more different uses than citizens (US=36; EU=30) and overall more uses were described in the US scenarios compared to the EU scenarios. Legal professionals focus more on AI tasks related to the operations of law firms. For example, none of the citizen scenarios address AI for case management, whereas six out of 25 legal expert scenarios in each of the US and the EU describe such use. Further, client interaction and support (e.g. consultation) was also more often described in expert scenarios. Comparing US and EU scenarios, our results show that document and content generation as well as legal research and analysis tasks were more prevalent in the US context.

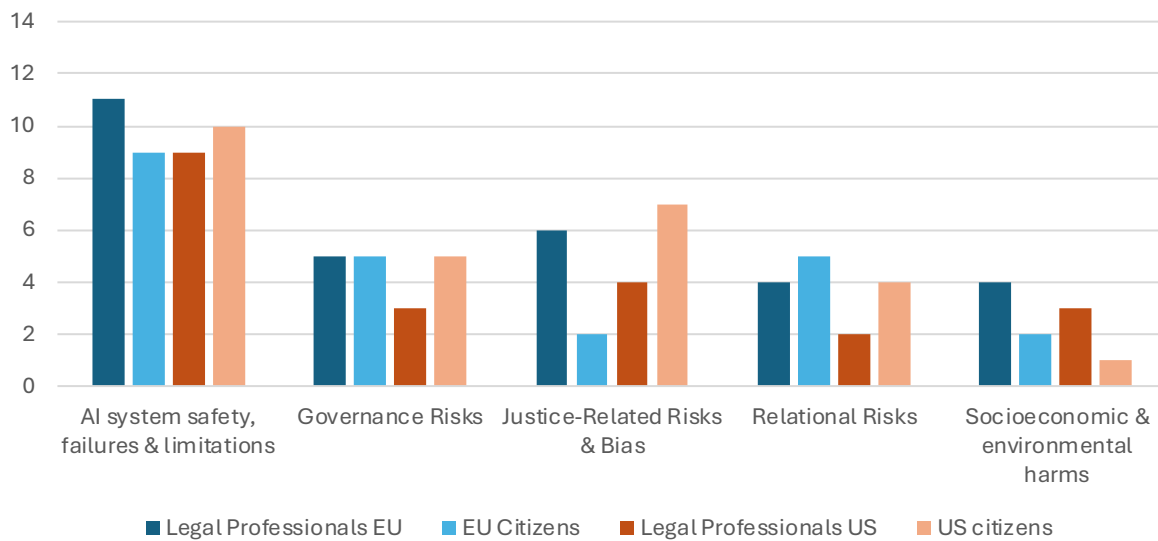

*Figure 2: Prevalence of Risks per Sample Group*

Regarding the variance in risk thematization (Figure 2), we see differences in the overall levels of risk mentions. The most critical group are EU legal professionals with 30 mentions of risks followed by US and EU citizens with 27 and 23 mentions. US legal professionals describe the fewest risks with only 21 mentions showing a stark contrast especially between the two legal professional groups. The results show that in all samples the most prevalent risks were AI system safety, failures & limitations, which most commonly were connected to a mention of AI errors of some forms (e.g. AI giving wrong advice). In general, there are only some secondary differences between the prevalence of risks between the different groups. For instance, socioeconomic harms were more prevalent in the legal professional samples; yet, these harms were also seldom mentioned at all. On the contrary, citizens (especially EU citizens) emphasize relational risks, which are commonly connected to the human element in the legal process. Further, especially US citizens are concerned about the impacts of GenAI on the quality of justice fearing biases and a decrease of justice. They share these concerns with EU legal professionals, whereas these risks are not common for EU citizens, and only marginally for legal professionals in the US.

Turning to the topic of benefits (see Figure 3), we see that EU citizens clearly diverge from the other groups in terms of total mentions of benefits as they only scarcely mention benefits in all (14 mentions int total in comparison to 23 for EU legal professionals, 27 for US citizens, and 25 for US legal professionals). A common denominator for all samples, however, is that costs and efficiency gains are consistently expected at an equal amount for all sample groups. Adding to the observable trends of the risks analysis, also the benefit analysis shows that US respondents tend to be more optimistic about the impacts of GenAI in the legal domain. The differences between the regional samples are especially evident for fairness and trust benefits, where US respondents describe more optimistic scenarios. Similarly, hopes are also put into the quality of output. Here, especially US citizens describe positive scenarios. In conjunction with the analysis of quality risks in the US sample, we also identify a polarization of the quality issue. While many scenarios from the US sample thematize quality of output, perceptions diverge in the sentiment of exactly that indicating that US stakeholders treat the quality of output as an important factor, but are yet unsure if GenAI will result in quality increase or decrease.

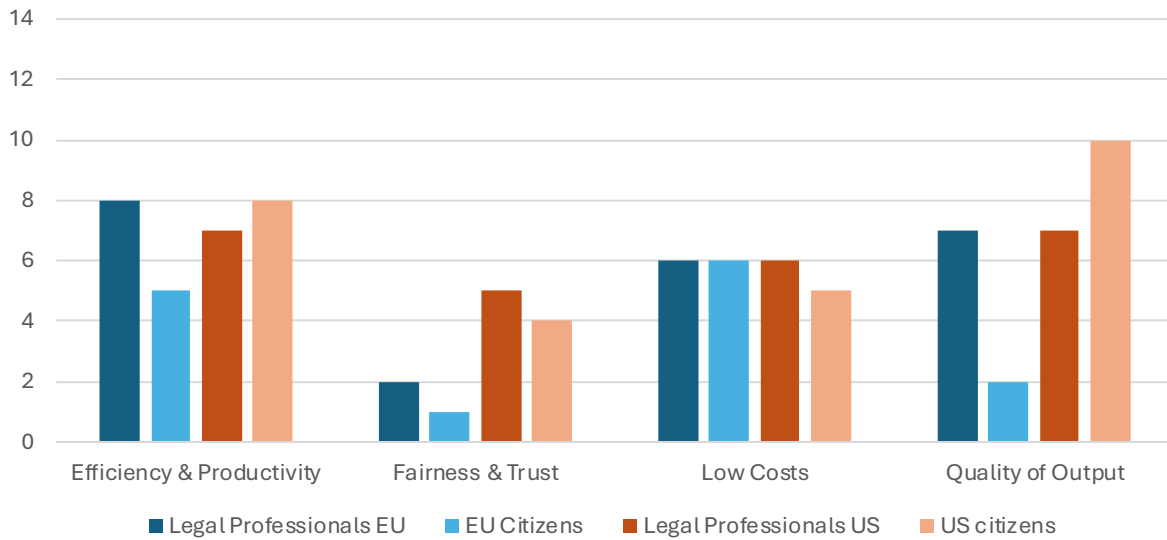

*Figure 3: Prevalence of Benefits per Sample Group*

### 5.5.    Risk and benefit associations regarding legal tasks

Next, we analysed the co-occurrence of legal tasks and risks and benefits mentions, i.e. which legal tasks are associated with which risk and benefits. In practice, we counted co-occurrences at the sub-section level. In other words, we identified the section of the scenario in which specific legal AI tools were described and counted the associated risks and benefits in that section. Thus, it was possible that multiple tools were mentioned in one scenario with different associated risks and benefits. Again, we only analysed co-occurrence on the aggregated main code level occasionally summarizing different legal tools (e.g. analysis of evidence and evidence collection on the main code level evidence and investigation). We do not differentiate between the different sample groups in this part of the analysis.

Our analysis (see Fig. 4 and 5) shows that the risk-benefit associations differed depending on the legal task for which AI was used. We also identified emerging trade-offs for tasks in which scenario writers described risks and benefits. These trade-offs are apparent for nearly every legal task described by scenario writers.

For case management and administration, the scenarios especially ascribed quality gains. The risks for these tasks are seldomly mentioned, but are distributed over all categories.

Client interaction and support benefits are foreseen as the quality of output increases (e.g. good legal advice), but also affordability is mentioned as an important benefit. In contrast, scenarios often describe the risks of AI failures (e.g. AI errors) as risk. On a secondary level, justice related risks & biases are mentioned, for instance that legal advice is biased and does not represent all positions. Hence, for client interaction and support we observe a quality trade-off. While some scenarios see potential in quality increases, others are concerned about exactly those promised quality gains. This hints towards a polarization regarding quality promises regarding the use of GenAI for legal tasks.

For decision making and adjudication tasks, benefits are mostly linked to efficiency and cost gains. It was often described that the use of AI helps relieving the burden of the costs as legal AI tools speed up the process and can even lead to avoiding court cases (e.g. through AI mediation). On the other side, respondents are concerned about potential AI errors that can be made in automated adjudication processes, which are also associated with justice-related risk (e.g. biased decision-making). Interestingly, in decision-making and adjudication tasks there are less mentions of quality gains through GenAI. Thus, we observe an efficiency and costs vs quality trade-off of legal decision making, so a high risk – high gain trade-off.

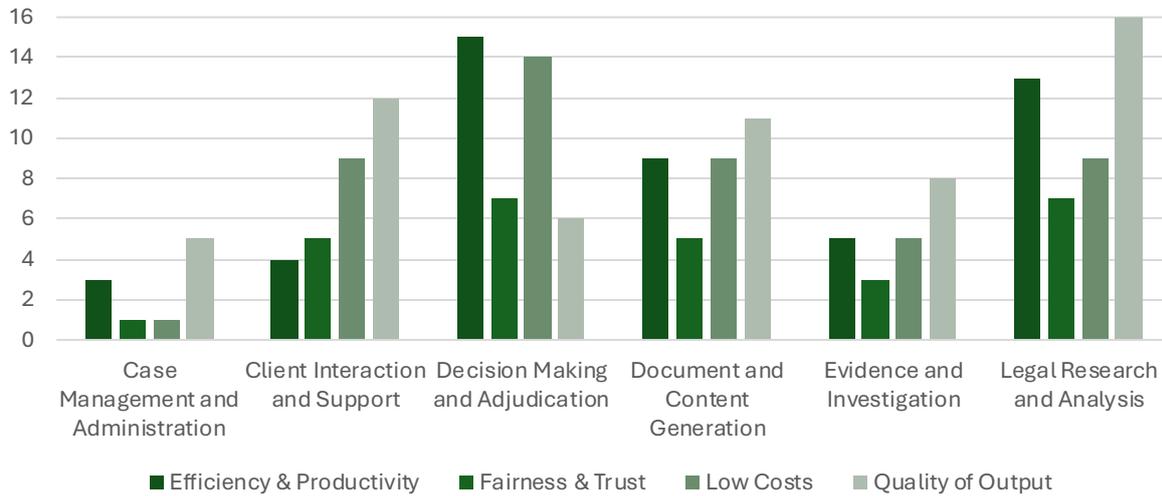

*Figure 4: Co-Occurrence of Legal Tasks and Benefits*

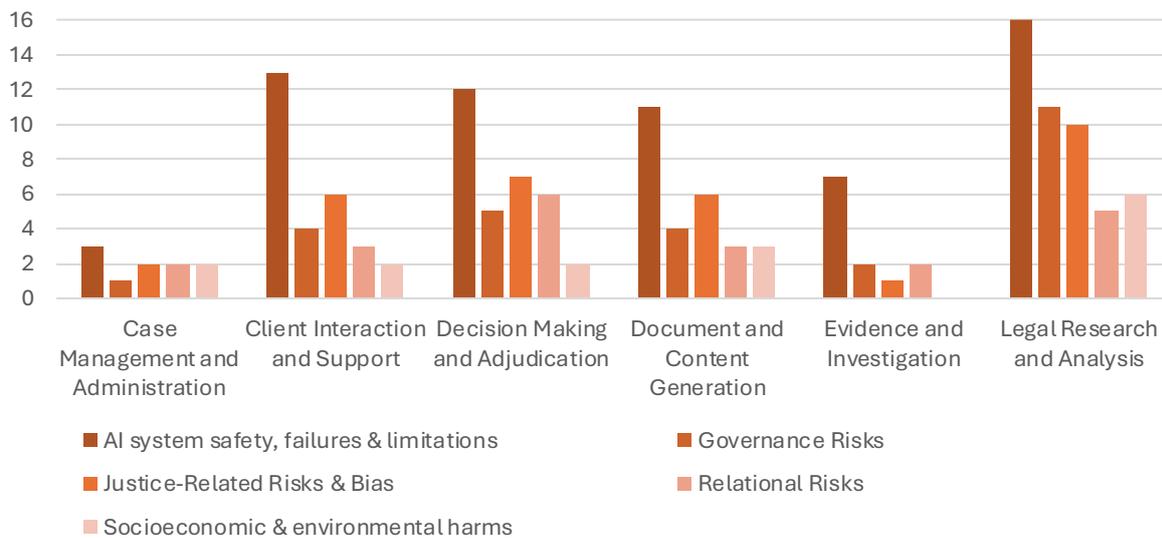

*Figure 5: Co-Occurrence of Legal Tasks and Risks*

Document and content generation as well as evidence and investigation tasks were mostly associated with quality gains, but also with efficiency and costs benefits. Similarly to most of the other tasks, risks were mostly seen in terms of the limitations of AI also hinting towards quality trade-offs. This is similar to the observations for the client interaction and support tasks.

The use of GenAI for legal research and analysis was also associated with quality and efficiency gains. Scenarios described how GenAI can search case law or legal data bases and provide a comprehensive overview in short time. On the other hand, some scenarios were concerned about exactly this quality of legal research. Scenarios described that legal research – while appearing to be comprehensive – lacked important information or interpreted laws wrong. Further, for this task, governance risks were especially apparent. These describe, for instance, the interdependence of AI systems (e.g. different tools giving different advice on the same case) or the societal acceptance of GenAI for legal research (e.g. if the use of GenAI is not acknowledged in court or by colleagues).

## 6. Discussion

This paper set out to map plausible future applications of GenAI in the legal sector in consideration of their potential risks and benefits. This is an increasingly important issue, as AI is set to be used more frequently in legal conflict resolution processes, which could change not only individual case outcomes but also the institution of justice as a whole (Helberger, 2025). In order for this process to be facilitated in the public interest, exploratory insights are provided that highlight potentially prevalent risk and gain areas. The study also identified plausible trade-offs that need to be navigated, as well as different perspectives, including the anticipations of legal professionals whose work routines will most likely be changed by AI, and citizens who are potential clients and conflict parties on the receiving end of a legal conflict. Our study also aims to help regulators, risk assessors, and other scholars refine risk assessment standards and define migration strategies aimed at safeguarding the digital transformation of the legal sector in an ethical manner (Pullen et al., 2026; Socol De La Osa & Remolina, 2024).

### 6.1. Adding a Qualitative, Context-Aware Element to Current Risk Assessment Practices

Our findings show that written narratives provide rich insights that can inform risk assessments. First, we identified a variety of anticipated uses of AI that could affect legal conflict resolution. GenAI was perceived to potentially transform legal conflict resolution along the entire value chain, from getting an initial overview of the legal case to drafting documents to automated adjudication. Although scenarios are not a predictive method by nature and are no guarantee whatsoever that future developments will materialise in exactly the way imagined, our research does suggest that some uses are relatively uncontroversial (e.g. use of GenAI in case management) whereas other use cases are clearly high-risk high-gain scenarios, like the use of GenAi in legal research or in decisions making and adjudication. These two uses happen to be also the applications where most cost and efficiency gains are expected, making them more likely cases for adoption, while also triggering most concerns.

The anticipated uses, risks and benefits also create ample of food for thought for the ongoing GenAI-transformation of the legal sector. For example, consider the use of GenAI for evidence collection that adds to existing professional debates on navigating AI evidence authentication (Runyon, 2025):

> "However, there is now the possibility of using AI to create another witness statement. To do this, the AI takes all possible existing evidence and can create its own testimony based on the audio recordings, images of the accident site and vehicles, and witness statements, and can precisely reconstruct the behavior of the accident using a video file". [EU_LP_18]

This examples opens up further questions such as: how should courts handle artificially generated statements? Should they be treated as trustworthy sources, equal to human witnesses? Or are they dismissed as irrelevant fictional statements? This is only one example of a discussion that could be held in the near future. Considering these questions now will enable decision-makers to react more quickly and influence the trajectory of technology development early on (Selbst, 2021).

Furthermore, we demonstrated that different perspectives are important for envisioning future applications of AI and for identifying its risks and benefits. As the different trends between legal professional and citizen scenarios as well as the regional contexts indicate, the outcome of risk assessments can differ greatly, depending on whose perspective is taken. Put differently, there is a real risk that generic risk assessments invisibilize (significant) differences in risk perceptions and the degree to which different stakeholders are affected by one and the same technological solution. Citizens highlight and evaluate future uses, concerns, or benefits that legal professionals and technology developers have not prioritized or identified (e.g. concerns about the vagueness of GenAI output). Our study further shows that citizens may be more concerned about automated decision-making, even if it is not the most plausible future use. These narratives can nevertheless influence perceptions of AI use and should be considered when aligning GenAI to the public interest.

Further, our study support recent literature in observing that regional context matters when it comes to perceptions of AI (N. Gillespie et al., 2025; Kelley et al., 2021). For instance, EU citizens tend to be more critical in their evaluation of legal AI tools than their US counterparts. This is important given the global reach of many technology companies. Citizens from different countries and regulatory spheres have different expectations towards technology that shape their risk anticipations and may be differently addressed by policy and regulatory approaches. Concretely this could mean, for example, that tools such as Anthropic's Legal Plugin are potentially more likely to be deployed more widely in the US than in the EU, at least for the time being. Drawing on the work of other researchers in the field (Gellert, 2020; Pasquale, 2023; Walker et al., 2024), we have also advocated for the inclusion of qualitative, tangible elements in risk assessment practices. We propose incorporating narratives that contain rich contextual information. This allows us to do more than list risks and benefits; it enables us to describe how affected parties experience them. This is particularly important for legal conflict resolution, as human motivations, perceptions, and narratives play a significant role in resolving such conflicts. Not only legal conflicts but also political decision-making depend on AI narratives. In these contexts, justifications, lived experiences, and perceptions could influence regulators in drafting and adapting legislation (Berresheim, 2024; Mager & Katzenbach, 2021).

## 6.2.    Critical Re-Assessment of Current Approaches to Risked-Based Governance

The insights from this study, while qualitative and explorative in nature, do raise a number of questions that can an inform a critical appraisal of existing risked-based approaches to AI governance in the legal sector. The challenge of any risked-based approach to AI governance is deciding which uses of AI are high or low risk (Gellert, 2020; Kieslich et al., 2025; Orwat et al., 2024). The EU AI Act, for example, draws the distinction at the use of AI by judicial authorities in researching, interpreting and applying the law to facts, as well as the use of dispute resolution bodies. The study calls into question the adequacy of qualifying the use of AI by judicial authorities high risk, while qualifying the use of (sometimes the same) technology by lawyers or citizens also in the area of dispute resolution as not risky. Examples are the AI safety concerns in client interaction and support, governance risks in case management and administration or around the use of GenAI to produce legal documents that both, professional users and citizen anticipate for the use of GenAI in legal conflict resolution. The study further questions the adequacy of a 'one-size-fits-all' approach to risk assessments, as risks can potentially differ quite significantly depending on and between the envisaged target groups (here professional users and citizens). A challenge for future risk assessment exercises, for example in the context of the European AI Act, is acknowledging and balancing differing risk perceptions and levels of acceptability.

The study also suggests that potential risks can stem not only from the technology itself, but also the way it is implemented and employed. In particular, respondents pointed to the lack of independence, lack of (internal) accountability but also power asymmetries as potential sources of risks that a strong technology-centric risk-assessment approach is not able to reveal. Neither are these risks that can be mitigated through technology design, but instead require risk mitigation measures in the organisational sphere. Finally, a contribution of this study, and the method in more general, is to help prioritise uses of AI that are potentially more risky than others (e.g. use in legal research vs case management), even if the specific prevalence findings we present shouldn't be generalized given the non-representative and exploratory nature of our samples. Seeing the many different potential applications of AI in the legal context and the limited resources of regulators as well as developers, having methods to prioritise risk areas and create some sort of 'risk heat map' can be an important contribution to the overall feasibility of the risked-based approach (Meßmer & Degeling, 2023; Orwat et al., 2024).

These findings are also relevant to the US context, where no comprehensive regulatory framework is established. Our results indicate that AI in the legal sector might plausibly cause serious risks that need to be managed in order to benefit society as a whole, whether or not comprehensive legal regulation is in place. While the establishment of AI regulations under the current administration seems highly unlikely, we nevertheless highlight the importance of our findings for standardization bodes like NIST. Those bodies, based on the results of our study, might consider enhancing standards to explicitly incorporate participation from both professional domain experts and from the public at large into risk identification and management processes.

### 6.3. Emerging Trade-Offs for AI Use in Legal Conflict Resolution

Beyond identifying potential risks, an important step for regulators and risk assessors is conducting risk management, i.e., deciding on acceptable risk levels. Most of these decisions involve weighing of risks and benefits (van der Heijden, 2021); and, in the bigger picture follow a thought that Helberger (2025) puts forward: "It is good to remember, however, that we do have a choice as long as we do not give up asking: who do we want to define the public values and human rights for the digital world?" (p.9). For appropriate trade-off decisions to be made collectively as a society, we argue that the trade-offs must first be explicated before it is possible to decide on the (non)acceptability of certain risks. Our study sheds light on some plausible trade-offs that risk assessors may need to make in the future. Note that, given the limited sample size and qualitative nature of the study, we do not claim that our list of risk-benefit trade-offs is exhaustive.

A consistent trade-off apparent in multiple AI applications is that between efficiency gains and system failures. AI systems promise to simplify various tasks, such as drafting documents or determining the outcome of legal conflicts. However, these efficiency promises are accompanied by concerns about the quality of the AI tool, as mistakes could have severe consequences for those affected. This raises the question of whether the core benefits of an AI transformation should be cost savings and efficiency, especially in the legal sector, and how potential cost and efficiency benefits for one party to the detriment of another should affect the distribution of liability. As one respondent pointed out in the scenario: "should it be allowed to delegate delicate calculations to an AI without human supervision? Is the ruling more likely if a human or a machine calculates? And in this case, who should be held responsible when the AI makes a mistake?"
[EU_C_15].

Quality trade-offs are also apparent for many legal tasks. While many scenarios outline positive quality expectations towards the use of AI, for instance, in giving helpful advice that ultimately benefits their users (e.g. winning a court case), other scenario writers are concerned about exactly those quality promises. Scenarios frequently describe how GenAI models make errors that lead to detrimental outcomes for users (e.g. losing a job in case of the use by legal professionals or losing a court case in case of citizens). This quality trade-off indicates uncertainty among legal professionals and citizens alike regarding the quality promises of AI – throughout most of the legal tasks that are imagined. One possible implication from this could be the need to rethink existing legal and judicial procedures and safeguards of legal decision making to enable the contestability of the use of GenAI in the legal sector. Traditionally, making decisions contestable, e.g. before court or a court of a higher instances, has served the goal of ensuring the 'quality' of legal decision making in terms of fairness, rule of law, adherence to professional standards and fundamental rights. The question is then whether and how the introduction of GenAI into legal conflict resolution might trigger the need to rethink these contestability mechanisms, for example through explainability rights, new standards of due process and professional integrity, but also standards that allow non-technical experts to understand limits and capabilities of models. This might, for example, take the form of new standard benchmarks of legal reasoning capability (e.g., evaluation standards of GenAI accuracy or argument coherence), or pragmatic metrics of legal AI success rates, though these would in turn raise new issues about just access and understanding of those ratings.

A related trade-off is between cost and model performance, particularly for client interaction tasks. Today, we can already observe people turning to GenAI for support with nearly all queries. The obvious benefit is lower costs, particularly for basic models like ChatGPT. However, studies have shown that the performance of basic AI tools and specialized legal tools varies (Magesh et al., 2024; Ryan & Hardie, 2024). Therefore, while access to information may be inexpensive, the reliability of the advice cannot be taken for granted. Ultimately, this issue comes down to how to balance these trade-offs. While one could argue that inexpensive AI tools provide easier access to justice, particularly for those with limited financial resources, the effectiveness of this access depends on the performance of the system and potential costs that may materialize later (e.g. if a party occurs a legal disadvantage or financial harm as a result). This access may either provide benefits or exacerbate existing inequalities. We emphasize the need for risk assessors to consider these trade-offs, taking into account the viewpoints of different affected groups.

Our findings further underscore observations regarding the influence on AI on the legal sector as Helberger (2025) points out: "These inroads of technology companies into the sphere of courts and justice are often accompanied by a rhetoric of urgency and inevitability if justice, as an institution, is hoping to survive." (p.5). Efficiency and quality gains are presented as one of the most urgent issues necessitating the use of AI. However, with this push towards justice as a service, they also implement their own values and may cause conflicts due to differing understandings of how justice should be served. In the worst case, this could lead to a power conflict between the legal system and technology companies that aim to implement AI tools for the legal sector, opening up a debate about whose values should be followed. As this discussion outlines, these larger questions also apply to individuals, such as legal professionals and citizens, who encounter value conflicts when interacting with technology. They contribute an important source of knowledge by translating researchers' abstract ideas into concrete, tangible examples of value conflicts.

### 6.4.    Limitations

Because our findings are qualitative, we can only point out trends in the data rather than perform statistical tests to detect significant differences. Therefore, our findings should be interpreted as exploratory indicators of future trajectories rather than as statistical evidence for describing probabilistic futures. Additionally, our European sample consisted of citizens of several different European countries. Although we analysed our data in the context of a joint European regulation, we acknowledge the heterogeneity of European countries.

## 7.  Conclusion

The legal sector is already experiencing the impact of GenAI, and researchers argue that these impacts may change legal institutions and professional values (Helberger, 2025). We argue that additional methods are needed to explore future technological developments and their associated risk-benefit trade-offs from a diversity of perspectives and incorporate more contextual information into risk management. Our findings demonstrate that scenario writing is an effective tool for activating projections not only from legal professionals but also from ordinary citizens. We describe various risk-benefit trade-offs that could arise in the future from using GenAI to resolve legal conflicts. Risk assessors and citizens must decide how the future of legal conflicts will evolve. We content that making informed trade-off decisions between efficiency and human intuition, costs and thoroughness, and the "neutrality" of AI versus biased outcomes will ultimately decide about the broader societal acceptability of the GenAI-assisted transformation of the legal sector. To do so, risk management decisions should be based not only on metrics but also on human experiences and hopes for the future.

**Funding**: The funding for this research was provided by UL Research Institutes through the Center for Advancing Safety of Machine Intelligence.

**Conflicts of interest/Competing interests:** The authors declare no competing interests.

**Ethical approval and informed consent statements**: We received ethics approval from the institutional ethics board of the first authors' affiliation.

## References

*AlgoSoc AI Opinion Monitor*. (2025). https://monitor.algosoc.org/values.html

Amer, M., Daim, T. U., & Jetter, A. (2013). A review of scenario planning. *Futures : The Journal of Policy, Planning and Futures Studies*, *46*, 23–40. https://doi.org/10.1016/j.futures.2012.10.003

Barnett, J., Kieslich, K., Sinchai, J., & Diakopoulos, N. (2025). Scenarios in Computing Research: A Systematic Review of the Use of Scenario Methods for Exploring the Future of Computing Technologies in Society. *arXiv Preprint arXiv:2506.05605*.

Berresheim, L. H. M. (2024). *The right to privacy* [Universiteit van Amsterdam]. https://pure.uva.nl/ws/files/197406328/Thesis.pdf

Börjeson, L., Höjer, M., Dreborg, K.-H., Ekvall, T., & Finnveden, G. (2006). Scenario types and techniques: Towards a user's guide. *Futures*, *38*(7), 723–739. https://doi.org/10.1016/j.futures.2005.12.002

Brauner, P., Glawe, F., Liehner, G. L., Vervier, L., & Ziefle, M. (2025). Mapping public perception of artificial intelligence: Expectations, risk–benefit tradeoffs, and value as determinants for societal acceptance. *Technological Forecasting and Social Change*, *220*, 124304. https://doi.org/10.1016/j.techfore.2025.124304

Chien, C. V., & Kim, M. (2024). Generative AI and Legal Aid: Results from a Field Study and 100 Use Cases to Bridge the Access to Justice Gap. *Loyola of Los Angeles Law Review*. https://ssrn.com/abstract=4733061

Chien, C. V., Kim, M., Raj, A., & Rathish, R. (2024). How Generative AI Can Help Address the Access to Justice Gap Through the Courts. *Loyola of Los Angeles Law Review, Forthcoming*. https://ssrn.com/abstract=4683309

Choi, J. H., Monahan, A., & Schwarcz, D. B. (2023). Lawyering in the Age of Artificial Intelligence. *SSRN Electronic Journal*. https://doi.org/10.2139/ssrn.4626276

De Sousa, W. G., Fidelis, R. A., De Souza Bermejo, P. H., Da Silva Gonçalo, A. G., & De Souza Melo, B. (2022). Artificial intelligence and speedy trial in the judiciary: Myth, reality or need? A case study in the Brazilian Supreme Court (STF). *Government Information Quarterly*, *39*(1), 101660. https://doi.org/10.1016/j.giq.2021.101660

De Vries, E., Schoonvelde, M., & Schumacher, G. (2018). No Longer Lost in Translation: Evidence that Google Translate Works for Comparative Bag-of-Words Text Applications. *Political Analysis*, *26*(4), 417–430. https://doi.org/10.1017/pan.2018.26

Dhungel, A.-K. (2025). "This Verdict was Created with the Help of Generative AI...?" On the Use of Large Language Models by Judges. *Digital Government: Research and Practice*, *6*(1), 1–8. https://doi.org/10.1145/3696319

Dhungel, A.-K., & Beute, E. (2024). AI Systems in the Judiciary: Amicus Curiae? Interviews with Judges on Acceptance and Potential Use of Intelligent Algorithms. *ECIS 2024 Proceedings*, *7*. https://doi.org/https://aisel.aisnet.org/ecis2024/track06_humanaicollab/track06_humanaicollab/7

Ebers, M. (2024). Truly Risk-Based Regulation of Artificial Intelligence-How to Implement the EU's AI Act. *SSRN Electronic Journal*. https://doi.org/https://dx.doi.org/10.2139/ssrn.4870387

Efroni, Z. (2021). The digital services act: Risk-based regulation of online platforms. *Internet Policy Review*.

ELTA. (2023). *Legal Professionals & Generative AI. Global Survey 2023*. https://elta.org/generative-ai-global-report-2023/

European Commission. Directorate General for Justice and Consumers. (2025). *The 2025 EU justice scoreboard: Communication from the Commission to the European Parliament, the Council, the European Central Bank, the European Economic and Social Committee and the Committee of the Regions.* Publications Office. https://data.europa.eu/doi/10.2838/9524751

Fischer-Abaigar, U., Kern, C., Barda, N., & Kreuter, F. (2024). Bridging the gap: Towards an expanded toolkit for AI-driven decision-making in the public sector. *Government Information Quarterly*, *41*(4), 101976. https://doi.org/10.1016/j.giq.2024.101976

Gellert, R. (2020). *The Risk-Based Approach to Data Protection* (1st ed.). Oxford University PressOxford. https://doi.org/10.1093/oso/9780198837718.001.0001

Gillespie, N., Lockey, S., Ward, T., Macdade, A., & Hassed, G. (2025). *Trust, attitudes and use of artificial intelligence: A global study 2025* (p. 4974511 Bytes). The University of Melbourne. https://doi.org/10.26188/28822919

Gillespie, T. (2024). Generative AI and the politics of visibility. *Big Data & Society*, *11*(2), 20539517241252131. https://doi.org/10.1177/20539517241252131

Glaser, B., & Strauss, A. (2017). *Discovery of grounded theory: Strategies for qualitative research*. Routledge.

Griffin, R. (2024). What do we talk about when we talk about risk? Risk politics in the EU's Digital Services Act. *DSA Observatory*. https://dsa-observatory.eu/2024/07/31/what-do-we-talk-about-when-we-talk-about-risk-risk-politics-in-the-eus-digital-services-act/

Helberger, N. (2024). FutureNewsCorp, or how the AI Act changed the future of news. *Computer Law & Security Review*, *52*, 105915. https://doi.org/10.1016/j.clsr.2023.105915

Helberger, N. (2025). The rise of technology courts, or: How technology companies re-invent adjudication for a digital world. *Computer Law & Security Review*, *56*, 106118. https://doi.org/10.1016/j.clsr.2025.106118

Henry, J. (2024). *We Asked Every Am Law 100 Law Firm How They're Using Gen AI. Here's What We Learned.* (The American Lawyer). https://www.law.com/americanlawyer/2024/01/29/we-asked-every-am-law-100-firm-how-theyre-using-gen-ai-heres-what-we-learned/

Kelley, P. G., Yang, Y., Heldreth, C., Moessner, C., Sedley, A., Kramm, A., Newman, D. T., & Woodruff, A. (2021). Exciting, Useful, Worrying, Futuristic: Public Perception of Artificial Intelligence in 8 Countries. *Proceedings of the 2021 AAAI/ACM Conference on AI, Ethics, and Society*, 627–637. https://doi.org/10.1145/3461702.3462605

Kieslich, K., Diakopoulos, N., & Helberger, N. (2024). Anticipating impacts: Using large-scale scenario-writing to explore diverse implications of generative AI in the news environment. *AI and Ethics*. https://doi.org/10.1007/s43681-024-00497-4

Kieslich, K., Došenović, P., Starke, C., & Lünich, M. (2021). Artificial Intelligence in Journalism: How does the public perceive the impact of artificial intelligence on the future of journalism? *Meinungsmonitor Kuenstliche Intelligenz*, (4).

Kieslich, K., Helberger, N., & Diakopoulos, N. (2024). My Future with My Chatbot: A Scenario-Driven, User-Centric Approach to Anticipating AI Impacts. *The 2024 ACM Conference on Fairness, Accountability, and Transparency*, 2071–2085. https://doi.org/10.1145/3630106.3659026

Kieslich, K., Helberger, N., & Diakopoulos, N. (2025). *Scenario-Based Sociotechnical Envisioning (SSE): An Approach to Enhance Systemic Risk Assessments*. https://doi.org/10.31235/osf.io/ertsj_v1

Kieslich, K., Morosoli, S., Diakopoulos, N., & Helberger, N. (2026). *Trade-Offs in Deploying Legal AI: Insights from a Public Opinion Study to Guide AI Risk Management* (Version 1). arXiv. https://doi.org/10.48550/ARXIV.2602.09636

Kim, T., & Peng, W. (2024). Do we want AI judges? The acceptance of AI judges' judicial decision-making on moral foundations. *AI & SOCIETY*. https://doi.org/10.1007/s00146-024-02121-9

Laptev, V. A., & Feyzrakhmanova, D. R. (2024). Application of Artificial Intelligence in Justice: Current Trends and Future Prospects. *Human-Centric Intelligent Systems*, *4*(3), 394–405. https://doi.org/10.1007/s44230-024-00074-2

Lind, F., Eberl, J.-M., Eisele, O., Heidenreich, T., Galyga, S., & Boomgaarden, H. G. (2022). Building the Bridge: Topic Modeling for Comparative Research. *Communication Methods and Measures*, *16*(2), 96–114. https://doi.org/10.1080/19312458.2021.1965973

Lofland, J., Snow, D., Anderson, L., & Lofland, L. H. (2022). *Analyzing social settings: A guide to qualitative observation and analysis*. Waveland Press.

Mager, A., & Katzenbach, C. (2021). Future imaginaries in the making and governing of digital technology: Multiple, contested, commodified. *New Media & Society*, *23*(2), 223–236. https://doi.org/10.1177/1461444820929321

Magesh, V., Surani, F., Dahl, M., Suzgun, M., Manning, C. D., & Ho, D. E. (2024). *Hallucination-Free? Assessing the Reliability of Leading AI Legal Research Tools* (Version 1). arXiv. https://doi.org/10.48550/ARXIV.2405.20362

Martinho, A. (2024). Surveying Judges about artificial intelligence: Profession, judicial adjudication, and legal principles. *AI & SOCIETY*. https://doi.org/10.1007/s00146-024-01869-4

Meßmer, A.-K., & Degeling, M. (2023). *Auditing Recommender Systems. Putting the DSA into practice wit a risk-scenario-based approach*. Stiftung Neue Verantwortung. https://www.stiftung-nv.de/sites/default/files/auditing.recommender.systems.pdf

Metikos, L. (2024). Explaining and Contesting Judicial Profiling Systems. *Technology and Regulation*, *2024*, 188–208. https://doi.org/10.71265/azacz070

National Institute of Standards and Technology. (2023). *Artificial Intelligence Risk Management Framework (AI RMF 1.0)* (NIST AI 100-1; p. NIST AI 100-1). National Institute of Standards and Technology (U.S.). https://doi.org/10.6028/NIST.AI.100-1

National Institute of Standards and Technology. (2024). *Artificial Intelligence Risk Management Framework: Generative Artificial Intelligence Profile* (NIST AI NIST AI 600-1; p. NIST AI NIST AI 600-1). National Institute of Standards and Technology. https://doi.org/10.6028/NIST.AI.600-1

Nazareno, L., & Douglas-Glenn, N. E. (2025). Six Years of Proposed AI Legislation Across the US States: What's on Policymakers' Minds?  *SSRN Electronic Journal*. https://doi.org/10.2139/ssrn.5782782

Orwat, C., Bareis, J., Folberth, A., Jahnel, J., & Wadephul, C. (2024). Normative Challenges of Risk Regulation of Artificial Intelligence. *NanoEthics*, *18*(2), 11. https://doi.org/10.1007/s11569-024-00454-9

Pasquale, F. (2023). Power and Knowledge in Policy Evaluation: From Managing Budgets to Analyzing Scenarios. *Law and Contemporary Problems*, *86*(3).

Pullen, E., Ruijer, E., & Meijer, A. (2026). Governing Ethics for the Digital Transformation: Developing, Testing, and Validating a Framework. *Government Information Quarterly*, *43*(2). https://doi.org/https://doi.org/10.1016/j.giq.2026.102117

Ramirez, R., & Wilkinson, A. (2016). *Strategic reframing: The Oxford scenario planning approach*. Oxford University Press.

Runyon, N. (2025). Deepfakes on trial: How judges are navigating AI evidence authentication. *Thomson Reuters*. https://www.thomsonreuters.com/en-us/posts/ai-in-courts/deepfakes-evidence-authentication/

Ryan, F., & Hardie, L. (2024). ChatGPT, I have a Legal Question? The Impact of Generative AI Tools on Law Clinics and Access to Justice. *International Journal of Clinical Legal Education*, *31*(1), 166–205. https://doi.org/10.19164/ijcle.v31i1.1401

Schmitz, A., Mock, M., Görge, R., Cremers, A. B., & Poretschkin, M. (2024). A global scale comparison of risk aggregation in AI assessment frameworks. *AI and Ethics*. https://doi.org/10.1007/s43681-024-00479-6

Schwarcz, D., Manning, S., Barry, P., Cleveland, D. R., & Rich, B. (2025). *AI-POWERED LAWYERING: AI REASONING MODELS, RETRIEVAL AUGMENTED GENERATION, AND THE FUTURE OF LEGAL PRACTICE*. https://papers.ssrn.com/sol3/papers.cfm?abstract_id=5162111

Selbst, A. D. (2021). AN INSTITUTIONAL VIEW OF ALGORITHMIC IMPACT. *Harvard Journal of Law & Technology*, *35*(1).

Slattery, P., Saeri, A. K., Grundy, E. A. C., Graham, J., Noetel, M., Uuk, R., Dao, J., Soroush Pour, Casper, S., & Thompson, N. (2024). *The AI Risk Repository: A Comprehensive Meta-Review, Database, and Taxonomy of Risks From Artificial Intelligence*. https://doi.org/10.13140/RG.2.2.28850.00968

Socol De La Osa, D. U., & Remolina, N. (2024). Artificial intelligence at the bench: Legal and ethical challenges of informing—or misinforming—judicial decision-making through generative AI. *Data & Policy*, *6*, e59. https://doi.org/10.1017/dap.2024.53

Stahl, B. C., Antoniou, J., Bhalla, N., Brooks, L., Jansen, P., Lindqvist, B., Kirichenko, A., Marchal, S., Rodrigues, R., Santiago, N., Warso, Z., & Wright, D. (2023). A systematic review of artificial intelligence impact assessments. *Artificial Intelligence Review*. https://doi.org/10.1007/s10462-023-10420-8

The Hague Institute for Innovation of Law, & The Institute for the Advancement of the American Legal System. (2021). *Justice Needs and Satisfaction in the United States of America 2021*. https://iaals.du.edu/sites/default/files/documents/publications/justice-needs-and-satisfaction-us.pdf

Tian, E., & Cui, A. (2023). *GPTZero: Towards detection of AI-generated text using zero-shot and supervised methods*. GPTZero. https://gptzero.me

Ullstein, C., Jarvers, S., Hohendanner, M., Papakyriakopoulos, O., & Grossklags, J. (2025). Participatory AI and the EU AI Act. *Proceedings of the AAAI/ACM Conference on AI, Ethics, and Society*, *8*(3), 2550–2562. https://doi.org/10.1609/aies.v8i3.36737

van der Heijden, J. (2021). Risk as an approach to regulatory governance: An evidence synthesis and research agenda. *Sage Open*, *11*(3), 21582440211032202.

Villasenor, J. (2024). Generative Artificial Intelligence and the Practice of Law: Impact, Opportunities, and Risks. *Minnesota Journal of Law, Science & Technology*, *25*. https://scholarship.law.umn.edu/mjlst/vol25/iss2/8

Walker, H., Pope, J., Morrison-Saunders, A., Bond, A., Diduck, A. P., Sinclair, A. J., Middel, B., & Retief, F. (2024). Identifying and promoting qualitative methods for impact assessment. *Impact Assessment and Project Appraisal*, *42*(3), 294–305. https://doi.org/10.1080/14615517.2024.2369454

## Appendix

**Appendix A: Instructions for scenario exercise**

**Please read the following instructions here and on the next pages very carefully as you need this information to perform your writing task.**
*Please note that the "Continue" button will only appear after a minimum amount of time.*

**Task Description**

In this exercise we ask you to write a short (minimum 300 words) **fictional scenario** to explore how the **generative AI technology** could affect **legal conflict resolution** in five years from now.

A **scenario** is a **short fictional story** that includes: 1) a **setting** of time and place, 2) **characters** with particular motivations and goals, and 3) a **plot** that includes character actions and **events** that lead to some impact of interest.

Generative AI refers to a technology that can create new content (e.g. text, images, audio, video) based on a prompt, instruction, and content the model was trained on, and any other content the model is provided with or has access to.

A **legal conflict** refers to a dispute between two or more parties that involves a **disagreement over legal rights, obligations, or interpretations of the law**. These conflicts can arise in various areas of law, including civil, criminal, business, family, or international law. Legal conflicts are typically resolved through **discussion, mediation, or in court.** Generative AI could be used in **all stages of legal conflict resolution** (e.g., formulating a complaint, information gathering, analysis of the law, document drafting, up to judicial decision-making).

**Specific Instructions:**

- Please develop your scenario **based on your professional perspectives, experiences, and knowledge.**
- Please develop your scenario to be **creative** and **original** in regard to the **impacts of generative AI on legal conflict resolution**.
- Please choose your **setting** to be five years in the future and in the society where you currently reside.

On the next pages you will see more details on the technology for the scenario we ask you to write. Please take your time to familiarize yourself with the information.

**Appendix B: Technology Description**

**Technology**

Generative AI is a technology that can create new content based on the content it was trained on.

**Capabilities**

- Generative AI can be used to rewrite, summarize, personalize, translate, or extract data based on input texts. It can also be set up as a chatbot that end-users can interactively communicate with.
- These AI systems can be controlled using text-based prompts which provide task instructions and input data. For instance, you could prompt it with:
    o "Summarize the following: <text>"
    o "Give legal advice for the following <issue>"
    o "Explain <issue> in easy to understand language"
    o "Create an official / legal document containing the following information <text>"
    o "Provide information about <issue>"

**Limitations**

- **Accuracy**: This technology does not always output text or information that is accurate. It can make mistakes.
- **Privacy and Security**: This technology processes sensitive personal information of users.
- **Biases**: The outputs from this technology can be biased based on the data used to train the system, which typically reflects common societal biases (e.g. racial or gender).
- **Lack of Contextual Understanding**: The technology may not understand the context of a question and/or user's intentions.

**Trends**

- Some Generative AI tools (e.g., ChatGPT; DeepSeek; Copilot) are freely available for users, i.e. everyone with an internet connection can use them.
- Some companies develop specialized legal generative AI tools that are targeted at lawyers and/or law companies.
- Generative AI has already been used in the legal sector for …
    o Counseling on legal documents (e.g., assist with document review, legal research memos, contract analysis)
    o Conducting legal research (e.g., search of relevant case law)
    o Supporting legal interpretations
    o Answering questions about the law
    o Predicting the outcome of cases
    o Preparing summaries and analysis of legal documents
    o Writing legal complaints

**Appendix C**

*C1: Example story from a EU legal professional [EU_LP_01]*

In 2030, artificial intelligence will be a routine part of everyday legal business. Law offices, courts and consumers will be using generative AI routinely to draft legal contracts, research case law, and provide legal counselling to clients. The government has enacted a regulatory regime to ensure compliance with AI legal services, but the pace of the technology's advance continues to raise ethical as well as legal questions. the following is a scenario of such a case.

Rudolf, a software programmer, has sued microsoftnova company, a global corporation, for breaching a software development contract. Rudolf has agreed to develop a proprietary AI algorithm for microsoftnova, but the company allegedly utilized the software without reasonably compensating him. Unable to afford services of a highly profiled lawyer, Rudolf relies on legalbot-499, a generative AI to assist with legal research, draft pleadings and even predict case outcomes based on precedents.

Sam, for microsoftnova, seeks to dismiss Rudolf's AI generated arguments on grounds that generative AI cannot replace human legal reasoning and that application of the AI may impose bias and inaccuracy. Judge turner, judge of the case in court, is aware of the growing use of AI bus is wary of its limitations , namely ethical responsibility and data bias.

As the trial unfolds, legalbot-499 finds a provision in the agreement that had been noticed previously and enhances the argument of Rudolf. Sam opposes by saying that arguments presented through AI cannot be admitted unless proved by a human lawyer. How can robots be used to judge cases or scenarios, yet they can't reason as a human being. It is left for the court to decide if legal arguments generated through AI are of equal credibility compared to arguments derived from human advocates. The AI would simply be replacing humans in their work.

This case is a milestone in determining the role of generative AI in determining legal differences. The ruling becomes a precedent for the acceptability of AI in legal practice, and how subsequent legislations written for AI assisted legal practice. Though AI is demonstrated to be beneficial to level the playing field for these who cannot afford expensive legal consultation, concerns over responsibility, accuracy and ethics are foremost for legal academics. Is the work of lawyers and legal professions in danger of being taken by generative AIs? This question can only be answered with time.

***C2: Example story from a EU citizen [EU_C_10]***

Back in 2030, using artificial intelligence to make legal decisions was completely normal for us. As we know, Portuguese courts have never been good at bureaucracy and cases used to fall apart because they took too long to process, but that's how DINIS came about. DINIS is a system that was developed by the big law firms in Portugal to automate their work, since more work equals more money - having a tool that made it easier meant that they made double or triple the same money with half the work.

One day, Joana, a lawyer who had just entered the profession, used DINIS to help her manage her workload and this day was particularly serious. Joana had been handed a delicate case involving allegations of digital abuse among teenagers, namely the sharing of sensitive images generated by artificial intelligence which added to the black humor of the situation. As the timeframe for dealing with this case was sensitive and, in fact, Joana had never had much practice with thinking for herself during her law degree, due to the constant use of platforms such as ChatGPT and DeepSeek, she asked DINIS to create an initial petition based on the case data and the latest case law.

Obviously, as all these platforms do, the document looked impeccable. It was well written and had solid legal references... or at least it seemed to. It was inevitable that DINIS would get it wrong and it turned out that one of the legal references he cited... was completely invented and Joana never noticed because she relied too much on this technology.

There was no avoiding what happened and, unluckily, Joana's client ended up incarcerated because his case wasn't strong enough due to the poor assistance of DINIS and Joana. It was only after the latter's insistence from inside prison and with the support of his family that another judge decided to take a deeper look at the case and realize the inconsistencies of the situation. Joana was obviously punished by the Bar Association for her mistake, and her firm was in bad shape.

This wasn't a surprising situation for anyone - these cases happened day in, day out with DINIS, but for some reason, it ended up making enough headlines and upsetting enough people to demand more legislation and more control over DINIS. It was also discovered that the platform was trained with internal documents from the big Portuguese firms that helped develop it and this meant that the tool was unintentionally biased in terms of having better legal strategies for cases where clients had more resources, making clients with fewer resources have quicker decisions and less legal assistance so that they were no longer a "dead weight" for the firms or a potential expense.

This discovery ended up generating a huge debate in Portugal, which had never happened before, about the use of artificial intelligence in law and in 2030, it became mandatory for all legal documents generated by systems such as DINIS to be reviewed by humans with specialization in the area of law and, above all, for the training data from this tool to be urgently audited. But until this happens, many lawyers, like Joana, have seen their reputations and years of dedication to the field of law tarnished by relying on non-human technology.

***C3: Example story from a US legal professional [US_LP_15]***

Doug and his wife Sharon had not been getting along. Doug had been considering divorce but he had heard horror stories from his friends and co-workers and he was afraid of losing half of his pension. Doug was also worried about Sharon taking full custody of the children, of which the parties had three under 18, if he tried to fight her too much on the financial end of the divorce. Doug set up a meeting with an attorney, and in his free consultation he discussed some of his legal concerns related to his desire for a divorce. He wasn't happy with the information that he heard from the attorney; namely, that giving up half of his pension was going to be expected of him. While he was aware that he had been married for over 20 years, he considered that money his own and did not believe it was fair for him to have to share his pension with his wife. Even worse, the attorney told him that he would likely have to pay some sort of spousal maintenance to Sharon. He did not like either of these pieces of information, and he also did not like the $15,000 retainer that he was quoted from the attorney. He decided to see what he could do on his own.

Doug began researching online to see if he could find answers that he felt were more applicable to his situation and his desired outcome for his divorce. When he ran an online search for "divorce" and "pension," he found lots of websites for attorneys discussing various ways that splitting a pension can be avoided, but almost all of them made reference to his wife getting at least part of his pension as being an inevitability. Doug began to have a suspicion that any attorney that he hired would get some sort of benefit out of having him split his pension with his wife Sharon although he couldn't figure out what that benefit might be. He decided that he would be better researching and making legal arguments on his own, so he could make his position known to the judge directly. Doug believed that if he could show the judge that Sharon had not contributed at all to the pension that he wouldn't have to split it with her. And so Doug decided that he wanted to try to use AI to teach him about the law and to help him to represent himself.

When he searched for "AI" and "divorce papers," Doug found multiple generative AI options available to him. He chose to go with chat GPT because it was the only one that he had heard of before. He started off by asking Chat GPT to help him draft divorce papers. He knew enough to at least say what state he was in and chat GPT helped him to generate a divorce petition after asking him some follow-up questions. Those questions included the date of his marriage, the names and ages of his children, and whether the parties had any agreements as to custody, property settlement, or asset and debt distribution. Doug didn't really understand the initial process for filing for divorce and so he stated the things that he wanted to be true, which were that he was entitled to split custody of his children, that Sharon was not entitled to any of his pension, and that the parties had decided to split the debt 50/50. There were a lot more questions in this process than he had anticipated, such as what he was going to do with the house, the bank accounts, and if he needed to make any arrangements for additional schooling for children in the future; specifically college expenses. In the end, chat GPT helped Doug to generate a petition that he thought presented his arguments well and he decided to move forward with it. He filed the papers along with the filing fee and his wife Sharon was served two days later.

Unlike Doug, Sharon went to a local attorney and decided that this was the route that she wanted to go. She found a female attorney who helped her to feel less confused about the process and spoke to her about some options as far as payment of the retainer went and gave her honest information about what Sharon's could expect during the divorce process. Sharon's attorney told her that Doug's petition was indicative of someone who did not know what they were doing, and that he was making all kinds of arguments and statements in there that simply were not true, legally or factually speaking, and that such statements were not permitted per court rules. Sharon's attorney filed a motion to have the petition dismissed for failure to meet court rules and regulations. In addition to including false claims about agreements between the parties, the petition also failed to meet local rules. While chat GPT was able to formulate the petition to match Illinois law (the state where the parties were located), it did not account for rules from the local jurisdiction or rules from the local Court that had to be included when parties were drafting and filing petitions.

At their first Court date, Sharon's attorney was able to get Doug's petition dismissed, although Doug had the right to refile. The judge encouraged Doug to find an attorney, a real human attorney, with whom Doug could discuss the law and his wishes and who could formulate a petition which would match rules and serve his needs. This outcome solidified Doug's belief that he would not get a fair shake in court, as he believes that the judge's ruling was unfair. Doug continued to use chat GPT in his court filing processes, much to the frustration of the judge and Sharon and her attorney. In the end Doug did not get what he was wanting out of the divorce and ended up having to pay Sharon's attorney which he thought was extremely unfair, in addition to having to split his pension.

While the overall outcome of the divorce was not necessarily unfair to Doug under the circumstances, legally speaking, it left him feeling extremely embittered and disadvantaged by the process. Rather than believing this impact was due to his choice of using chat GPT as opposed to an attorney, he believed that the court and the entire judicial process was prejudiced against him because of his gender. Doug and Sharon lost a lot of time that they could have been trying to work together due to Doug's bitterness and Sharon's frustration with the excessive time and energy that was having to be spent trying to get Doug to hire a attorney. Because Doug was in charge of his own research, he was subject to his own biases as far as what he was seeing and what he wanted to see. As a result, the information that he fed into the chat GPT process was not objective and did not give chat GPT a standard basis from which to begin drafting Doug's court documents and presenting Doug with legal strategies.

***C4: Example story from a US citizen [US_C_04]***

In the near future of 2030, Mark had gone out to pick up his friend to surprise him for his birthday party. Traffic was faster than usual this day, and upon reaching an overpass mark crashes into a sudden accident. After regaining his composure, the police arrive on the scene. It turns out there had been debris left on the road from an unsecured truck, and caused the cars ahead to come to a stop as they crashed, Mark being the last impact.

After everything had been settled, nobody injured, and the roads cleared of debris, Mark returned home and opened a common generative AI he would use to ask questions to. He began to ask who would be at fault in the scenario he had just experienced, and after some back and forth with adding what context he could remember, was told that in that case he would not be at fault. Mark sighed with relief as he is currently going through financial hardship. One thing Mark failed to mention was that he had been changing insurance companies, and at the time of the accident he was uninsured. Soon enough Mark received a call from one of the police officers from the accident, and was told that he could ignore the ticket, as they had found the truck that left the debris. At this point, Mark was reassured, and found himself confident in the information the generative AI had told him.

A couple months later, Mark receives a summons from a debt collection agency that was looking to settle money owed for the damages one of the others from the accident received. Unknown to Mark who relied on legal advice and information given to him by the generative AI, due to his lack of insurance, there was no settlement between the other persons insurance company and his own. Suddenly, Mark was demanded $20,000 in damages, which bewildered him as he had believed he was not at fault. At this point he'd realize that though legal issues may be relayed to a context based text generator, there are many other factors that might pose a threat to his legal and financial safety. In a panic, Mark begins to type to the generative AI, asking questions regarding his options, legal advice, how these things happen, and even began asking it to scour the internet for information regarding similar cases. At this point, though he ends up more informed, he realizes at the point he's in now, the most he can do is go to a mediation and try to settle this with the debt collectors that the insurance company offloaded the debt to. Due to the turmoil and back and forth, Marks confidence in his understanding of the legal situation spirals, as he realizes that no matter how much he asks for information regarding what he could do the generative AI has limitations in what it can say, as without context understanding the many variables and intentions of the different parties involved in this situation is near impossible.

After settling for monthly payments that Mark can barely afford, he had learned more about his limited understanding of legal issues, and the legal world in general. Though information is so readily available to use, comprehension and critical thinking can find itself scarce when not exercised with intent. Convenience is king, but once without, one can find themselves incapable.